\documentclass[
    aps,
    prr,
    superscriptaddress,
    twocolumn,
    a4paper,
    floatfix,
    longbibliography
]{revtex4-2}

\usepackage[american]{babel}
\usepackage[utf8x]{inputenc}

\usepackage{grffile} 

\usepackage{newtxtext,newtxmath}
\usepackage{microtype}

\usepackage{amsmath}
\usepackage{amssymb}
\usepackage{bbm}
\usepackage{mathtools}
\usepackage{dsfont}
\usepackage{braket}
\usepackage{cancel}
\usepackage{slashed}

\usepackage{graphicx}
\usepackage[svgnames]{xcolor}

\usepackage{physics}

\usepackage{siunitx} 

\usepackage{ragged2e}
\usepackage{array}
\usepackage{tabularx}
\usepackage{tabulary}
\usepackage{booktabs}

\definecolor{cset-aps-blueberry}{RGB}{28,128,158}
\definecolor{cset-aps-blue}{RGB}{46,44,184}
\definecolor{cset-aps-turquoise}{RGB}{0,67,88}
\definecolor{cset-aps-limegreen}{RGB}{190,219,67}
\definecolor{cset-aps-green}{RGB}{31,138,112}
\definecolor{cset-aps-yellow}{RGB}{255,225,25}
\definecolor{cset-aps-orange}{RGB}{253,116,0}
\definecolor{cset-aps-red}{RGB}{219,0,43}

\usepackage{tikz}

\usepackage{pgfplots}
\pgfplotsset{%
    every axis legend/.append style={%
        cells={anchor=west},
        at={(0.96,0.04)},
        anchor=south east,
        font=\scriptsize,
        },
    every axis/.append style={%
        yticklabel style={%
            /pgf/number format/fixed zerofill,
            /pgf/number format/precision=2},
        },
    width= \textwidth,
    height=8cm,
    xmajorgrids=true,
    xminorgrids=false,
    minor x tick num=1,
}
\usepgfplotslibrary{external}

\usetikzlibrary{decorations}
\usetikzlibrary{decorations.pathmorphing}
\usetikzlibrary{decorations.pathreplacing}

\usepackage{pict2e,picture}

\makeatletter
\DeclareRobustCommand{\Arrow}[1][]{%
\check@mathfonts
\if\relax\detokenize{#1}\relax
\settowidth{\dimen@}{$\m@th\rightarrow$}%
\else
\setlength{\dimen@}{#1}%
\fi
\sbox\z@{\usefont{U}{lasy}{m}{n}\symbol{41}}%
\begin{picture}(\dimen@,\ht\z@)
\roundcap
\put(\dimexpr\dimen@-.7\wd\z@,0){\usebox\z@}
\put(0,\fontdimen22\textfont2){\line(1,0){\dimen@}}
\end{picture}%
}
\makeatother

\usepackage{hyperref}
\hypersetup{%
    colorlinks=true,
    linkcolor={cset-aps-red},
    linkbordercolor={cset-aps-red},
    filecolor={cset-aps-orange},
    filebordercolor={cset-aps-orange},
    citecolor={cset-aps-blue},
    citebordercolor={cset-aps-blue},
    urlcolor={cset-aps-green},
    urlbordercolor={cset-aps-green},
    menucolor={cset-aps-limegreen},
    menubordercolor={cset-aps-limegreen},
    breaklinks=true,
    pdfborderstyle={/S/U/W 2},
    pdfpagemode=UseOutlines,
    pdfstartpage={1},
}

\newcommand{\ee}{\text{e}}
\newcommand{\ii}{\text{i}}

\usepackage{lipsum}

\newcommand{\affHAN}{\address{Institut f{\"u}r Quantenoptik, Leibniz Universit{\"a}t Hannover, Welfengarten 1, D-30167 Hannover, Germany}}
\newcommand{\affULM}{\address{Institut f{\"u}r Quantenphysik and Center for Integrated Quantum Science and Technology (IQ\textsuperscript{ST}), Universit{\"a}t Ulm, Albert-Einstein-Allee 11, D-89069 Ulm, Germany}}
\newcommand{\affTUDa}{\address{Institut f{\"u}r Angewandte Physik, Technische Universit{\"a}t Darmstadt, Schlossgartenstr. 7, D-64289 Darmstadt, Germany}}

\begin{document}

\title[Light-pulse atom interferometry with entangled atom-optical elements]{Light-pulse atom interferometry with entangled atom-optical elements\vspace{.25em}}
\collaboration{This article has been published in \href{https://doi.org/10.1103/PhysRevResearch.4.013115}{Physical Review Research \textbf{4}, 013115 (2022)} by the American Physical Society under the terms of the \href{https://creativecommons.org/licenses/by/4.0/}{Creative Commons Attribution 4.0 International} license.}
\newcommand{\orcid}[1]{\href{https://orcid.org/#1}{\includegraphics[width=7pt,height=7pt]{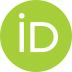}}}
\author{Tobias Asano\,\orcid{0000-0002-6257-8815}}
\email{tobias.asano@alumni.uni-ulm.de}
\email{tobias-asano@outlook.de}
\affULM
\author{Fabio Di Pumpo\,\orcid{0000-0002-6304-6183}}
\affULM
\author{Enno Giese\,\orcid{0000-0002-1126-6352}\,}
\affTUDa
\affHAN

\begin{abstract}
The analogs of optical elements in light-pulse atom interferometers are generated from the interaction of matter waves with light fields.
As such, these fields possess quantum properties, which fundamentally lead to a reduced visibility in the observed interference.
This loss is a consequence of the encoded information about the atom's path.
However, the quantum nature of the atom-optical elements also gives an additional degree of freedom to reduce such effects:
We demonstrate that entanglement between all light fields can be used to erase information about the atom's path and by that to partially recover the visibility.
Thus, our work highlights the role of complementarity on atom-interferometric experiments.
\end{abstract}

\maketitle
\section{introduction}
Light-pulse atom interferometery~\cite{Kasevich1991,Storey1994,Tino2014} is a powerful tool with unique applications~\cite{Bongs2019}, such as high-precision gravimeters~\cite{McGuirk2002}, gyroscopes~\cite{Gustavson1997}, and tests of fundamental physics~\cite{Schlippert2014,Parker2018,Morel2020,DiPumpo2021}.
These interferometers are implemented in a semi-classical manner, where beam splitters and mirrors are realized by diffraction from intense classical light pulses~\cite{Rasel1995}.
In light of the current drive towards optical cavity-based atom interferometers~\cite{Hamilton2015,Canuel2018,Xu2019,Nourshagh2020, Nourshagh2020_2}, we study the effect of quantized light fields generating the atom-optical elements~\cite{Soukup2021}, and we discuss scenarios in which they are entangled with each other.

Atomic diffraction from optical fields in cavities depends on the particular photon statistics of the light~\cite{Meystre1989, Akulin1991, Herkommer1992}.
Therefore, the quantum nature of light can be used as a lever to control the diffraction and even perform quantum operations on the center-of-mass (c.m.) motion of the atom~\cite{Khan1998}.
During the diffraction process, both the atom and the light field become entangled.
As a consequence, the atom can be used for quantum nondemolition measurements on the light field~\cite{Holland1991,Khan1999} and there are schemes to reconstruct entanglement between multiple cavities~\cite{Khosa2004}.
In turn, there have been proposals to generate quantum states of light through atom interferometers~\cite{Islam2008}.

Besides these single-atom considerations, Raman superradiant transitions~\cite{Schneble2004,Yoshikawa2004} or diffraction from optical cavities with quantized light fields are one promising route to generate quantum states with metrological gain for atom interferometry~\cite{Davis2016,Hosten2016}.
The sensitivity of atom interferometers can be enhanced even further if one performs measurements on the diffracting light fields~\cite{Haine2013,Haine2015,Haine2016, Greve2021}, thus recycling the information.

However, there is also a downside to the entanglement between atom and light, and the quantized nature of light can be detrimental for metrological applications~\cite{Nolan2017}.
For example, information about an atom's path may be encoded into light fields and give rise to a significant drop in visibility.
Such effects on atom interferometers have already been studied~\cite{Soukup2021}, and superpositions are one possible route to overcoming such issues, where, as a prime example, intensive coherent states give rise to the classical limit.
However, another feature of quantum light has not been studied in this context so far: the possibility to entangle the diffracting light fields.

In our article, we study initial entanglement of atomic beam splitters and mirrors in a Mach-Zehnder interferometer to partially restore the loss of visibility arising from quantized pulses.
This way, we shed light on aspects of complementarity: the connection between a reduced visibility of the interference signal and the corresponding presence of full welcher-Weg (which-way) information~\cite{Rasel1995,Scully1991,Storey1994,Englert1995,Duerr1998} encoded into the light fields. 
While quantum eraser experiments~\cite{Scully1982,Walborn2002} overcome this obstacle by erasing the information after the measurement, we use initial entanglement to suppress the physical process of imprinting welcher-Weg information.

\section{Interferometer model}
We model the atomic diffraction used for the atom-optical manipulation via the light-matter interaction in a Jaynes-Cummings model~\cite{Jaynes1963, Cummings1965, Schleich2001}.
In this description, the Hamiltonian 
\begin{align} \label{eq:JCH}
    \hat{H} = \hbar \omega_a \ket{e}\bra{e} + \frac{\hat{p}^2}{2m} + \hbar \omega \hat{a}^\dagger \hat{a} + \frac{\hbar \Omega}{2}   \ee^{\ii k \hat{z}} \hat{a} \ket{e}\bra{g} + \mathrm{H.c.}
\end{align}
governs the system's dynamics where a quantized light field with frequency $\omega$ interacts with a two-level system, consisting of a ground state $\ket{g}$ and an excited state $\ket{e}$ separated by the energy difference $\hbar \omega_a$.
In addition to the conventional model, we include the atom's c.m. motion and add a kinetic term with the atom's momentum $\hat{p} $ and mass $m$.
The bosonic annihilation and creation operators $\hat{a}$ and $\hat{a}^\dagger$ of the light field in Eq.~\eqref{eq:JCH} obey the canonical commutation relation $\comm{\hat{a}}{\hat{a}^\dagger}=1$.
Moreover, the displacement operator $\exp{\pm \ii k \hat{z}}$ is responsible for the momentum transfer $\pm \hbar k$ upon an internal transition and contains the wave number $k$ of the light field.
It depends on the position $\hat{z}$ of the atom, obeying the commutator relation $[\hat{z},\hat{p}]= \ii \hbar$.
In particular, we associate the annihilation of a photon encoded in $\hat{a}$ with a momentum transfer $\hbar k$, while the creation of a photon from $\hat{a}^{\dagger}$ is associated with the momentum transfer $-\hbar k$, in accordance with energy-momentum conservation.
Here, $\Omega=\abs{\Omega} \ee^{\ii \theta}$ is the complex coupling constant of the light-matter interaction, and it can be separated in amplitude $\abs{\Omega}$ and phase $\theta$.
Solving the Schr\"odinger equation on resonance for this Hamiltonian, the time evolution can be written in terms of the scattering operator
\begin{align}
\label{eq:ScattMatrix}
\begin{split}
    \hat{\mathcal{S}} &\equiv  \hat{c}_{\hat{n}+1} \, \ket{e}\bra{e} -\ii \, \ee^{\ii \, \qty( k \, \hat{z} +\theta)} \, \hat{a} \, \frac{\hat{s}_{\hat{n}}}{\sqrt{\hat{n}}} \, \ket{e}\bra{g}  \\
    &-\ii \, \ee^{-\ii \, \qty(  k \,\hat{z} +\theta)} \, \frac{\hat{s}_{\hat{n}}}{\sqrt{\hat{n}}} \, \hat{a}^{\dagger} \, \ket{g}\bra{e} + \hat{c}_{\hat{n}} \, \ket{g}\bra{g}
\end{split}
\end{align}
for short pulses and monochromatic plane waves~\cite{Kleinert2015}, similar to Ref.~\cite{Soukup2021}. 
In addition, we disregard the effects of a detuning and velocity selectivity in the scattering operator.
Here, we use the abbreviations 
\begin{align}
        \hat{c}_{\hat{n}} \equiv \cos(\frac{\Theta}{2} \sqrt{\frac{\hat{n}}{\bar{n}}}) \quad\text{ and }\quad
        \hat{s}_{\hat{n}} \equiv \sin(\frac{\Theta}{2} \sqrt{\frac{\hat{n}}{\bar{n}}}).
\end{align}
Moreover, $\hat{n}= \hat{a}^\dagger \hat{a}$ denotes the photon-number operator, $\Theta = \abs{\Omega} \sqrt{\bar{n}} \tau$ is the pulse area for the pulse duration $\tau$, and $\bar{n}$ is the average photon number of the light field.
The scattering operator $\hat{\mathcal{S}}$ encodes resonant Rabi oscillations between the ground and excited state based on single-photon transitions. 
Depending on the pulse duration $\tau$ for a fixed average photon number $\bar{n}$ of the field, a pulse area of $\pi/2$ and $\pi$ realizes a beam splitter or mirror, respectively.
\begin{figure}
    \centering
    \includegraphics[width=1\columnwidth]{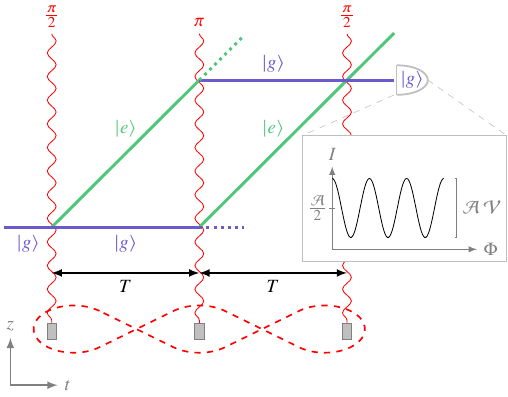}
    \caption{
    Spacetime diagram of a closed Mach-Zehnder atom interferometer based on single-photon transitions.
    A first $\pi / 2$ pulse creates an internal and c.m. superposition of the atom that enters the interferometer in the ground state $\ket{g}$ (blue).
    This superposition generates two branches.
    The excited state $\ket{e}$ (green) evolves for a time $T$ on the upper branch, whereas the ground state is associated with the lower branch.
    A $\pi$ pulse reverts the respective populations, and both branches are exposed to a time evolution for another time interval $T$. 
    They are then mixed by a final $\pi/2$ pulse giving rise to the interference signal $I$ depicted on the right side of the figure that can be measured by postselecting on the ground-state population.
    The interference fringes generated by the phase $\Phi$ can be observed with visibility $\mathcal{V}$ and amplitude $\mathcal{A}$, shown in the inset. 
    In general, imperfect pulses give rise to atom losses that decrease the detected signal and are highlighted by dotted lines.
    The dashed line circling the laser sources indicates an initial entanglement of the three light fields.
    Even though the figure shows single-photon transitions as also employed in the Bord\'e-interferometer~\cite{borde1989}, an analogous diagram for two-photon Raman diffraction gives rise to similar effects, as explained in the main body of the article.
    }
    \label{fig:1}
\end{figure}

Single-photon Rabi oscillations have been implemented for Rydberg atoms~\cite{Brune1996_2,Bertet2001} in cavity-based setups. 
These high-finesse cavities with small mode volumes are necessary to reach a sufficiently high interaction strength.
While it is possible to drive Rabi oscillations with low photon numbers, it is impractical for atom interferometry:
The sensitivity of accelerometers depends on the enclosed spacetime area, which is maximized in high-precision devices.
This drive toward large spatial separations~\cite{Kovachy2015} is in conflict with the requirements of small mode volumes.
Current developments of cavity-based atom interferometers~\cite{Hamilton2015,Canuel2018,Xu2019,Nourshagh2020, Nourshagh2020_2} strive for a cleaner mode structure and higher intensities.
However, there will be no application of such setups to a low-photon number regime in the foreseeable future.

Even though the first generation of light-pulse atom interferometers with traveling waves was based on single-photon transitions~\cite{borde1989, Riehle1991}, two-photon transitions inducing Raman or Bragg diffraction have developed into the state of the art~\cite{Kasevich1991,Torii2000}.
They have the benefit of working with long-lived ground states, where the duration of the interferometer and by that its sensitivity are not limited by severe restrictions given by the finite lifetime of the involved atomic states.
However, in light of proposals of detectors for gravitational waves and dark matter~\cite{Graham2013,Arvanitaki2018}, single-photon clock transitions as described by the Jaynes-Cummings model have again shifted into the focus of research~\cite{Hu2017, Hu2019, Rudolph2020}.
Yet, such schemes are limited by the lifetime of the excited state, in the case of a strontium clock transition~\cite{Rudolph2020} approximately to $0.14 \, \mathrm{ms}$.

Independent of this development, we emphasize that effective two-photon transitions could have been used in our model in a similar manner.
In this case, the photonic creation and annihilation operators are replaced by raising and lowering operators of the Schwinger representation of angular momenta~\cite{Schwinger1952} that encodes two radiation modes.
However, we focus in this article on single-photon transitions for clarity, but we emphasize that a generalization is possible but leads to a more cumbersome description.

Hence, by introducing the time evolution defined by
\begin{align}
\label{eq:U}
    \hat{U}\equiv \exp{-\ii\left(\frac{\hat{p}^2}{2m\hbar}+ \sum_{\ell=0}^2{ \omega \hat{n}_\ell} +\omega_a \ket{e}\bra{e}\right)T}
\end{align}
between two light pulses separated by a time interval $T$, we define the operator sequence that describes the evolution through the Mach-Zehnder interferometer shown in Fig.~\ref{fig:1} by
\begin{align}
\label{eq:OpSequence}
    \hat{U}_\text{MZ}\equiv \hat{\mathcal{S}}^{\qty(2)}\hat{U}\hat{\mathcal{S}}^{\qty(1)}\hat{U}\hat{\mathcal{S}}^{\qty(0)}.
\end{align}
The three pulses in the sequence of Eq.~\eqref{eq:OpSequence} denote individual solutions to the Schrödinger equation for the interaction of the $\ell$th light field with the atom. To this end, we add a superscript $\ell$ to the scattering operator from Eq.~\eqref{eq:ScattMatrix} for each pulse.
In contrast to the single pulse operator $\hat{\mathcal{S}}$, the $\ell$th pulse described by $\hat{\mathcal{S}}^{\qty(\ell)}$ contains modified operators $\hat{n}_\ell = \hat{a}^\dagger_\ell \hat{a}_\ell$ with $[\hat{a}_\ell,\hat{a}^\dagger_k]= \delta_{\ell,k}$, independent mean photon numbers $\bar{n}_\ell$ and pulse areas $\Theta_\ell$, as well as individual phases $\theta_\ell$ arising from the coupling constant.
Similar to conventional classical fields, we use pulse areas $\Theta_1=\pi$ and $\Theta_0=\Theta_2=\pi/2$, that is, beam splitters sandwiching a mirror pulse. 
The pulse area of the $\ell$th pulse $\Theta_\ell$ may be realized by either adjusting the mean photon number $\bar{n}_\ell$ or the pulse duration $\tau_\ell$, which can be chosen independently for every light field.
Note that the phase $\theta_\ell$ of the coupling constant might also differ from pulse to pulse.
Formally, the three pulses $\hat{\mathcal{S}}^{\qty(\ell)}$ act on field states $\ket{\psi_\ell}$ that reside in individual Hilbert spaces $\mathcal{H}_\ell$.

The observed interference signal of the interferometer depicted in Fig.~\ref{fig:1} takes the form
\begin{align}
\label{eq:IntSignalOp}
   I=\bra{\Psi} \otimes \bra{g} \hat{U}^\dagger_{\mathrm{MZ}} \hat{\Pi}_g \hat{U}_{\mathrm{MZ}} \ket{g} \otimes \ket{\Psi}= \frac{\mathcal{A}}{2} \qty(1+ \mathcal{V} \cos \Phi)
\end{align}
and results from a postselection on the ground state~\cite{Kleinert2015} through the projector $\hat{\Pi}_g=\op{g}$.
The expectation value is taken for an atom initially in the ground state $\ket{g}$ and an initial state $\ket{\Psi}$ containing the c.m. motion and the three light fields.

The transition element of the evolution operator that describes an atom entering and exiting the interferometer in the ground state takes the form $\ev{\hat{U}_{\mathrm{MZ}}}{g} = \hat{\mathcal{O}}_l + \hat{\mathcal{O}}_u$  and can be divided~\cite{Kleinert2015} into a superposition of two relevant branches, as shown in Fig.~\ref{fig:1} and derived in Appendix~\ref{sec:appendix}.
The operators $\hat{\mathcal{O}}_l$ and $\hat{\mathcal{O}}_u$ act on the c.m. motion of the atom and on the light fields, and they describe the propagation along the lower and upper branch.
They contribute to the interference signal from Eq.~\eqref{eq:IntSignalOp} through the visibility $\mathcal{V}$ and phase $\Phi$ defined by
\begin{align}
\label{eq:VisAndPhase}
    \mathcal{V} \ee^{\ii \Phi} = \frac{2 \big<\hat{\mathcal{O}}^{\dagger}_\text{l}\hat{\mathcal{O}}_\text{u} \big>_{\Psi}}{\big<\hat{\mathcal{O}}^{\dagger}_\text{l}\hat{\mathcal{O}}_\text{l}+\hat{\mathcal{O}}^{\dagger}_\text{u}\hat{\mathcal{O}}_\text{u}\big>_{\Psi}}
\end{align}
as well as the amplitude $\mathcal{A}\equiv2\big<\hat{\mathcal{O}}^{\dagger}_\text{l}\hat{\mathcal{O}}_\text{l}+\hat{\mathcal{O}}^{\dagger}_\text{u}\hat{\mathcal{O}}_\text{u}\big>_{\Psi}$.
Here, the index $\Psi$ underlines that all expectation values are taken with respect to $\ket{\Psi}$, which describes the initial c.m. motion and initial light fields.
We emphasize that $\mathcal{V}$ may flip its sign due to negative values of the trigonometric functions $\hat{s}_{\hat{n}_\ell}$ and $\hat{c}_{\hat{n}_\ell}$.
In contrast to that, we include phases encoded in the states of the light fields always in $\Phi$.
The three possible combinations of operators take the form
\begin{subequations}
\label{eq:SingleContribSignal}
\begin{align} 
    \label{eq:Ola}
    &\hat{\mathcal{O}}^{\dagger}_\text{u}\hat{\mathcal{O}}_\text{u}=\hat{c}^2_{\hat{n}_2}\otimes \hat{s}^2_{\hat{n}_1+1}\otimes \hat{s}^2_{\hat{n}_0}, \\ \label{eq:Olb}
    &\hat{\mathcal{O}}^{\dagger}_\text{l}\hat{\mathcal{O}}_\text{l}=\hat{s}^2_{\hat{n}_2+1}\otimes \hat{s}^2_{\hat{n}_1}\otimes \hat{c}^2_{\hat{n}_0},\\
    \label{eq:Olc}
    \mathrm{and} \quad& \hat{\mathcal{O}}^{\dagger}_\text{l}\hat{\mathcal{O}}_\text{u} = \ee^{\ii  \Delta \theta}  \hat{a}_2  \frac{\hat{s}_{\hat{n}_2} \hat{c}_{\hat{n}_2}}{\sqrt{\hat{n}_2}} \otimes \qty( \frac{\hat{s}_{\hat{n}_1}}{\sqrt{\hat{n}_1}}\hat{a}^\dagger_1)^2\otimes \hat{c}_{\hat{n}_0}\hat{a}_0  \frac{\hat{s}_{\hat{n}_0}}{\sqrt{\hat{n}_0}},
\end{align}
\end{subequations}
because the field and the atomic operators commute.
The c.m. motion cancels in Eqs.~\eqref{eq:Ola} and~\eqref{eq:Olb}.
However, in the overlap from Eq.~\eqref{eq:Olc} it reduces to a pure phase factor given by $\Delta \theta = \theta_0 -2 \theta_1 + \theta_2$ if the interferometer closes in phase space~\cite{Schleich2013}.
In the following discussion, it therefore suffices to solely consider light field states $\ket{\psi} \in \mathcal{H}_0 \otimes \mathcal{H}_1 \otimes \mathcal{H}_2$ when computing the expectation values in Eq.~\eqref{eq:VisAndPhase}, as the expectation values are independent of the initial c.m. state included in $\ket{\Psi}$.

For instance, light fields in a Fock state give rise to a vanishing visibility~\cite{Agarwal2003}, because they carry full welcher-Weg information of the atom, as discussed later in Fig.~\hyperref[fig:3]{\ref*{fig:3}(a)}. 
Contrary to coherent states whose phases can be inferred from the interference signal, atom interferometers cannot be used to infer the phase of Fock states.
In fact, Fock states lead to a vanishing visibility so that a phase measurement is meaningless.

In any case, the vanishing visibility can be partially compensated by involving superpositions of Fock states, whose phase difference can be inferred from the interference signal~\cite{Soukup2021}.
For example, a superposition of two Fock states in each field gives rise to a nonvanishing visibility that reaches $1 / 8$ for high photon numbers. 
Moreover, coherent states that are a good quantum-mechanical approximation for classical laser pulses lead to unit visibility for high photon numbers and introduce the known laser-phase contribution into the classical interference signal~\cite{Bertet2001,Soukup2021}.

\section{Classes of entanglement} \label{sec:2}
While there are established approaches to quantifying entanglement of multipartite two-level systems~\cite{Coffman2000, Bennett2000}, the situation becomes more subtle for the entanglement of multiple Fock spaces~\cite{Keyl2003,Duan2017}.
However, we select a subset of two orthogonal states $\ket{\downarrow_\ell}$ and $\ket{\uparrow_\ell}\in \mathcal{H}_\ell$ for each pulse $\ell = 0,1,2$.
In particular, we require orthogonality $\ip{\downarrow_\ell}{\uparrow_\ell}=0$ and normalization $\ip{\downarrow_\ell}{\downarrow_\ell}= \ip{\uparrow_\ell}{\uparrow_\ell}=1$.
Consequently, we restrict ourselves to fields
\begin{align} \label{eq:state}
    \ket{\psi} = \sum_{ijk} a_{ijk} \ket{i_0j_1k_2}
\end{align}
for $ijk \in \{\downarrow, \uparrow \}$ that can be expressed through these subsets.

To shed more light on different categories of tripartite entanglement, we briefly discuss two prime examples of entangled states:
The \emph{Greenberger-Horne-Zeilinger (GHZ) state}~\cite{Greenberger1989}
\begin{align} \label{eq:GHZ}
    \ket{\mathrm{GHZ}}= \frac{1}{\sqrt{2}} \qty( \ket{\downarrow_0,\downarrow_1,\downarrow_2} + \ket{\uparrow_0,\uparrow_1,\uparrow_2})
\end{align}
has the property of a separable reduced density operator $\hat{\rho}^{\qty(\ell)}_{\mathrm{GHZ}}=\Tr_\ell \ket{\mathrm{GHZ}}\bra{\mathrm{GHZ}}$ independent of the light field $\ell$ traced over. 
This type of entanglement is referred to as \textit{residual} entanglement. 
On the other hand, the \emph{W state}~\cite{Duerr2000} is defined as
\begin{align}
\label{eq:W-state}
    \ket{\mathrm{W}} = \frac{1}{\sqrt{3}} \qty( \ket{\downarrow_0, \downarrow_1, \uparrow_2} + \ket{\downarrow_0, \uparrow_1, \downarrow_2} + \ket{\uparrow_0,\downarrow_1,\downarrow_2}),
\end{align}
where the reduced density operator remains entangled for any partial trace over one light field. 
This class of entanglement is also known as \textit{pairwise} entanglement~\cite{Coffman2000}.

\begin{figure}[htb!]
    \centering
    \includegraphics[width=1\columnwidth]{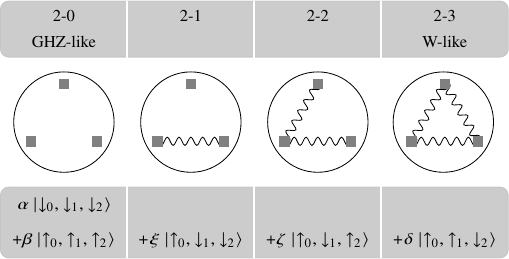}
    \caption{
    One possibility to categorize the entanglement of tripartite states.
    The four classes are denoted by 2-$j$, where the number $j$ indicates the amount of light fields that can be traced out of the tripartite system, so that the two-party system remains entangled.
    We identify the 2-0 class with GHZ-like states and the 2-3 class with W-like states.
    The circles around the three systems (denoted by squares) correspond to general tripartite entanglement, while the connection through wiggly lines between two parties represents pairwise entanglement.
    Example states for these classes are shown in the table below and are constructed by adding all components of the desired class. 
    For every class, all coefficients have to be nonvanishing except in the W-like class, where we can choose a vanishing $\beta$ and/or $\xi$.}
    \label{fig:entanglement}
\end{figure}

Both states are inseparable and thus share tripartite entanglement. 
However, they differ by the amount of light fields that can be traced out of the tripartite system, while the remaining two-party systems continue to be entangled.
This feature can be used to distinguish and quantify tripartite entanglement~\cite{Sabin2008}, as illustrated in Fig.~\ref{fig:entanglement}.
For example, states like the W state from Eq.~\eqref{eq:W-state} carry tripartite entanglement, denoted by the circle.
Tracing over any of the three subsystems (visualized by squares) leaves the other two subsystems pairwise entangled (symbolized by wiggly lines).
In contrast, tracing over any subsystem of a GHZ-like state leaves the other two separable (and thus not connected by wiggly lines).
We denote the different classes of states by 2-3 and 2-0 for three possible pairwise entangled combinations after the trace, or no combinations at all.

It is also possible to find situations in between:
When we consider a 2-1 state, that is, a GHZ-like state adding $\xi \ket{\uparrow_0,\downarrow_1,\downarrow_2}$, the reduced density operator remains entangled only if we trace out one light field, but a trace over the other two subsystems leads to a separable state.
An analogous scenario is shown in  Fig.~\ref{fig:entanglement} for the 2-2 state. 

Since we use a two-state subset, we can rely on established measures of tripartite entanglement.
Following Ref.~\cite{Coffman2000}, we introduce the measure
\begin{align} \label{eq:tau3}
    \tau_3 \equiv 2 \bigg| &\sum a_{ijk} a_{i'j'm} a_{npk'} a_{n'p'm'}  \epsilon_{ii'} \epsilon_{jj'} \epsilon_{kk'} \epsilon_{mm'} \epsilon_{nn'} \epsilon_{pp'} \bigg|,
\end{align}
where the sum is over all combinations of indices that take the values $\{\downarrow,\uparrow\}$.
Here, the Levi-Civita symbol $\epsilon_{ij}$ takes the values $\epsilon_{\downarrow \downarrow } = \epsilon_{\uparrow \uparrow} = 0$ and $\epsilon_{\downarrow \uparrow} = -\epsilon_{\uparrow \downarrow}=1$.
The measure $\tau_3 \in [0,1]$ grows with the amount of residual entanglement within the system, i.e., it is maximal for the GHZ state and minimal for the W state. The residual entanglement reduces to
\begin{equation} \label{eq:tau3spec}
    \tau_3 = 4 \abs{\alpha \beta }^2
\end{equation}
for all states defined in the table of Fig~\ref{fig:entanglement}.

\section{Initially entangled light fields}
In the previous section, we introduced different entangled states that differ by the amount of residual entanglement expressed through $\tau_3$.
In the following, we investigate how this difference influences the resulting visibility $\mathcal{V}$ of the interference signal. 

\subsection{Entangled Fock states}
As we solely require orthogonality and normalization of the two-state subset from Sec.~\ref{sec:2} that defines the state $\ket{\psi}$ of the three light fields, there are many possibilities in its specific choice.  
In the simplest case, we use Fock states $\ket{\downarrow_\ell} \equiv \ket{n_\ell}$ and $\ket{\uparrow_\ell} \equiv \ket{m_\ell}$ for all three light fields with photon numbers $n_\ell \neq m_\ell$.
We already saw that we can characterize entangled states according to their amount of residual entanglement.
Therefore, we start with the largest possible residual entanglement given by a GHZ-like state,
\begin{align} \label{eq:ghz-like}
\ket{\psi_{2-0}} = \alpha \, \ee^{\ii \vartheta/2} \, \ket{n_0,n_1,n_2} + \beta \, \ee^{-\ii  \vartheta/2} \, \ket{m_0,m_1,m_2}.
\end{align} 
Here, the real constants $\alpha$ and $\beta$ satisfy  $\alpha^2 + \beta^2 =1$, and $\vartheta$ denotes the phase difference between the two components.
While there are approaches to generate GHZ-entangled photonic states of a discrete or continuous nature, the fields have to interact with the atom at different times.
Hence, the three pulsed modes have to be delayed, e.g., through optical fibers and beam lines, or by using spatially separated cavities.
The latter requires an extension of our description by one dimension, adding a longitudinal velocity to the atomic beam~\cite{Berg2015} or relying on fountain-type experiments~\cite{Kovachy2015}.
However, the generation of three entangled and spatially separated modes or introducing a temporal delay comes with major issues like the deterioration of such difficile quantum states due to losses.
Producing such states is therefore challenging and beyond the scope of this article.

The overlap
\begin{align}\label{eq:ghz_vis}
\begin{split}
   \big<\hat{\mathcal{O}}^{\dagger}_\text{l}\hat{\mathcal{O}}_\text{u} \big>_{2-0} &=  \ee^{\ii \Phi } \alpha \beta \, c_{n_2}s_{n_2}
   s_{n_1+2}s_{n_1+1}c_{n_0-1}s_{n_0}  \propto \sqrt{\tau_3}/2
\end{split}
\end{align}
results, for example, from the choice $m_0 =n_0-1$, $m_1=n_1+2$, and $m_2=n_2-1$, and we replaced $\alpha \beta = \sqrt{\tau_3}/2$ from Eq.~\eqref{eq:tau3spec}.
In the interference signal, we observe the resulting interferometer phase
\begin{equation}\label{eq:interferometerphase}
    \Phi = \Delta\theta + \vartheta,
\end{equation}
where $\vartheta$ enters instead of the familiar laser phase~\cite{Storey1994,Schleich2013}.

\begin{figure*}
    \centering
    \includegraphics[width=\textwidth]{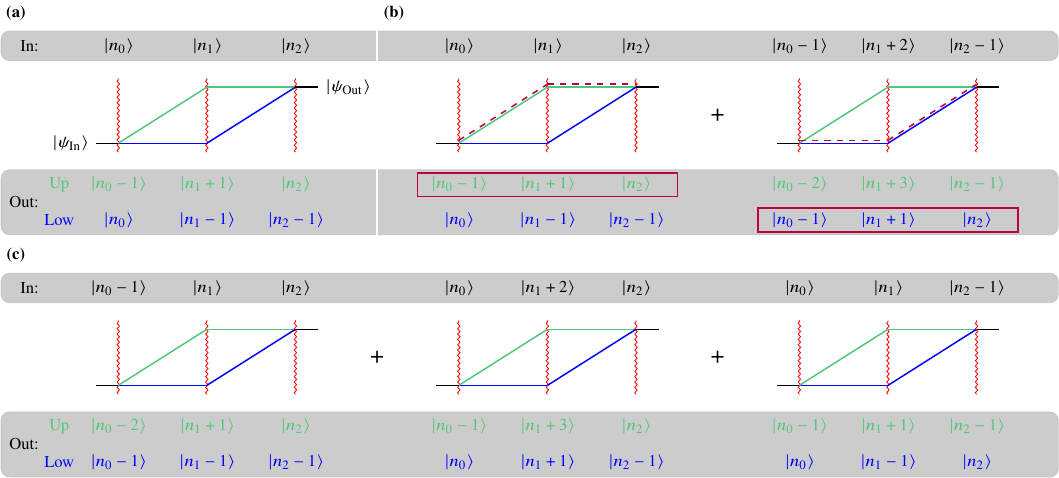}
    \caption{
    Modification of Fock states due to diffraction in a Mach-Zehnder interferometer.
    For separable Fock states (a), the evolutions along lower and upper branch are associated with unique photon numbers in all three fields.
    Measuring the state of any field unveils the atom's path, giving access to complete welcher-Weg information.
    A GHZ state (b) is a superposition of two such components.
    The photon numbers from the first component that resulted from a propagation along the upper branch are identical to the ones from the second  that arose from a propagation along the lower branch.
    The identical photon numbers are highlighted by purple boxes, and the respective propagation is denoted by purple dashed lines.
    Because they are indistinguishable, these components interfere and make it impossible to determine the atom's path with absolute certainty.
    For the example of a W state (c), we see that each possible combination of output photon numbers can be associated uniquely with either the upper or the lower branch.
    Hence, a measurement of all three fields gives access to the atom's path with absolute certainty.}
    \label{fig:3}
\end{figure*}

For $\beta=0$ ($\tau_3 =0$), the state from Eq.~\eqref{eq:ghz-like} reduces to separable Fock states in each light field, and we observe a vanishing visibility.
As illustrated in Fig.~\hyperref[fig:3]{\ref*{fig:3}(a)}, the photon numbers remaining after a propagation along the upper and lower branch differ.
Consequently, the light field contains full welcher-Weg information about the atom's path, wiping out the interference signal.

Conversely, for $\beta \neq 0$ we observe a nonvanishing visibility.
In particular, for $\alpha = \beta = 1/\sqrt{2}$ (i.e., $\tau_3=1$) and in the limit of high photon numbers, where the difference between $n_\ell$ and $n_\ell \pm 2$ becomes negligible~\cite{Soukup2021}, it takes the value
\begin{equation}
    \mathcal{V}= 1/2 > 0.
\end{equation}
Recovering the visibility through entanglement corresponds to erasing welcher-Weg information, as shown in Fig.~\hyperref[fig:3]{\ref*{fig:3}(b)}. 
The GHZ state is a superposition of two separable components.
For each component, one can infer individually full welcher-Weg information as described above.
However, the propagation caused by diffraction from two different components becomes indistinguishable.
For our specific choice of photon numbers, the photon numbers after the propagation along the upper branch of the first component and along the lower branch of the second component (highlighted by purple boxes) are identical.
Therefore, they lead to interfering contributions.
A trace over all fields results in only half of the cases in a distinct determination of the atom's path.
This fact explains the improvement of the visibility to $1/2$ but not to unity for high photon numbers, because there is still partial welcher-Weg information encoded into the light fields.

In contrast to the GHZ state, a W-like state,
\begin{align}
\begin{split}
\ket{\psi_{2-3}} = &\alpha  \ee^{\ii  \vartheta_1}   \ket{n_0,n_1,m_2} + \beta \ee^{\ii  \vartheta_2}  \ket{n_0, m_1,n_2}  \\ 
&+ \xi \ee^{\ii  \vartheta_3}  \ket{m_0,n_1,n_2},
\end{split}
\end{align}
is a superposition of three tripartite Fock states and contains the least amount of residual entanglement.
A pairwise comparison of these superposed states reveals that there is one photon number in common.
As the overlap operator in Eq.~\eqref{eq:Olc} contains annihilation and creation operators, the expectation value always vanishes, and we find
\begin{align}
\big<\hat{\mathcal{O}}^{\dagger}_\text{l}\hat{\mathcal{O}}_\text{u} \big>_{2-3} =0.
\end{align}
Even though the light fields were initially entangled, we still observe no interference signal.
The reason is illustrated in Fig.~\hyperref[fig:3]{\ref*{fig:3}(c)}:
We have again full welcher-Weg information about the atom's path, since a potential measurement of all three light fields uniquely reveals whether the atom has propagated along the upper or lower branch, respectively.

We have already discussed the visibility for the maximal (GHZ state with $\tau_3=1$) and the least amount (W state with $\tau_3=0$) of residual entanglement.
In between, we investigate the class 2-1 and 2-2 states defined according to the table in Fig.~\ref{fig:entanglement} with a symmetric choice of coefficients. 
\begin{figure}[htb]
    \centering
    \includegraphics[width=1\columnwidth]{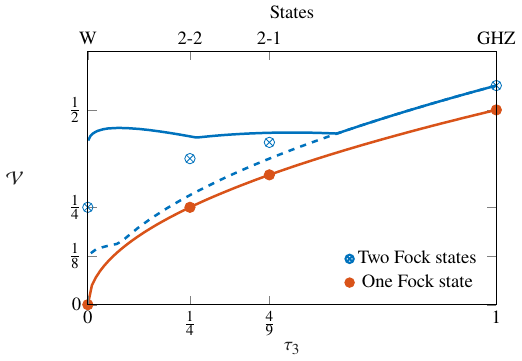}
    \caption{
    Visibility $\mathcal{V}$ in the high photon limit as a function of residual entanglement $\tau_3$. 
    We specify the two-state subset through Fock states (red) and observe $\mathcal{V}=\sqrt{\tau_3}/2$.
    The states with equal superpositions are denoted by dots and show that the GHZ state gives rise to the best visibility, whereas it vanishes for the W state.
    For superpositions of Fock states (blue) in each component of the entangled state, the GHZ state again displays the best visibility. 
    If we vary the coefficient $\alpha$ without changing Fock numbers, we find the blue dashed curve.
    The behavior is qualitatively similar to the results for one Fock state.
    However, for $\tau_3 \to 0$ the visibility approaches $1/8$ as expected for a separable superposition~\cite{Soukup2021}.
    Expanding the GHZ state by adding more components, selecting equal superpositions, and choosing optimal photon numbers, the 2-1, 2-2 and W states give rise to the visibility denoted by blue crosses.
    The blue curve corresponds to a continuous scan of $\tau_3$, i.e., not to an equal superposition.
    For each value of $\tau_3$, the combination of photon numbers was chosen to maximize the visibility.}
    \label{fig:result}
\end{figure}

We use Eq.~\eqref{eq:tau3spec} to determine the amount of residual entanglement for these states.
Similar to the W-like state, additional components in the states 2-1 and 2-2 cannot contribute to additional terms in the overlap so that we arrive again at Eq.~\eqref{eq:ghz_vis}.
Figure~\ref{fig:result} illustrates the visibility where the circles denote the results for symmetric coefficients.
Increasing the coefficients continuously while keeping the overall state normalized to unity corresponds to tuning $\tau_3$ and gives rise to the continuous line in Fig.~\ref{fig:result}.
For simplicity, the visibility is shown for values corresponding to the high photon limit, where we replace the trigonometric functions that arise, for example in Eq.~\eqref{eq:ghz_vis}, by the ideal values of $1/\sqrt{2}$ for a beam splitter pulse and by unity for the mirror pulse.
We see that the visibility rises as we increase the amount of residual entanglement within the states and reaches its maximum for the GHZ state.

Beyond the limit of high photon numbers, we can exploit the asymmetric property of beam splitters and mirrors in the regime of quantized light fields to maximize the visibility.
For instance, in Fig.~\hyperref[fig:3]{\ref*{fig:3}(b)}, we saw that only one path of each component is able to interfere.
Consequently, by increasing the amount of population in these branches, it is possible to attain a higher visibility than the presented results, in particular for low photon numbers~\cite{Soukup2021}.
However, since the overlap reduces to the same expression for all the states, the improvement is a scaling factor independent of the initial state, and the qualitative behavior of Fig.~\ref{fig:result} remains the same.

\subsection{Entangled superposed Fock states}
\label{sec:superpos}
We saw that entanglement can erase  welcher-Weg information under certain conditions.
A similar effect can be observed for superpositions of Fock states~\cite{Soukup2021}.
Therefore, we expect that a combination of superpositions and entanglement will improve the visibility even further.
To this end, we choose the subset
\begin{align} 
\label{eq:basis_superposition}
    \ket{\downarrow_i} \equiv   \frac{\ee^{\ii  \gamma_i}  \ket{m_i} + \ee^{\ii  \varepsilon_i }  \ket{n_i} }{\sqrt{2}} \quad\text{and}\quad
    \ket{\uparrow_i} \equiv \frac{\ee^{\ii \delta_i} \ket{r_i} + \ee^{\ii \eta_i }  \ket{s_i} }{\sqrt{2}}
\end{align}
that consists of a superposition of Fock states with photon numbers $m_i$, $n_i$, $r_i$, and $s_i$ that are pairwise different.
Moreover, we introduce the phases $\gamma_i, \varepsilon_i, \delta_i, \eta_i$ that we specify later. 

\begin{table}
\caption{
Two-state subsets consisting of superpositions of Fock states for the GHZ and W states.
The photon numbers and phases specified here are chosen such that we find nonvanishing overlaps and an effective phase $\vartheta=\vartheta_0+\vartheta_1+\vartheta_2$ observed in the interference signal.
\vspace{.5em}
}
    \begin{tabularx}{\columnwidth}{c
   >{\centering\arraybackslash}X 
   >{\centering\arraybackslash}X }
       \toprule[0.1pt] \toprule[0.1pt] \addlinespace
      State   & GHZ & W  \\ \addlinespace \midrule[0.1pt] \addlinespace
       $\ket{\downarrow_0}$  & $\frac{ \ket{n_0}+ \ee^{\ii 2  \vartheta_0}  \ket{n_0+2}}{\sqrt{2}}$ & $\frac{\ee^{\ii \vartheta_{0}/2} \ket{n_0} + \ee^{ -\ii  \vartheta_{0}/2}  \ket{n_0-1}}{\sqrt{2}}$ \\ \addlinespace
       $\ket{\uparrow_0}$  & $\frac{ \ket{n_0-1}+ \ee^{\ii  2   \vartheta_0}  \ket{n_0+1}}{\sqrt{2}}$ & $ \frac{\ee^{\ii 3 \vartheta_{0}/2} \ket{n_0+1} + \ee^{ -\ii 3\vartheta_{0}/2}  \ket{n_0-2}}{\sqrt{2}}$ \\ \addlinespace \midrule[0.1pt]
       $\ket{\downarrow_1}$  & $\frac{ \ket{n_1}+ \ee^{\ii  2 \vartheta_1}  \ket{n_1-4} }{\sqrt{2}}$ & $\frac{\ee^{\ii \vartheta_{1}/2} \ket{n_1}+ \ee^{ -\ii \vartheta_{1}/2}  \ket{n_1+2}}{\sqrt{2}}$ \\ \addlinespace
       $\ket{\uparrow_1}$  & $\frac{ \ket{n_1+2}+\ee^{\ii  2  \vartheta_1}  \ket{n_1-2}}{\sqrt{2}}$ & $\frac{\ee^{\ii 3\vartheta_{1}/2} \ket{n_1-2}+ \ee^{ -\ii 3\vartheta_{1}/2}  \ket{n_1+4}}{\sqrt{2}}$ \\ \addlinespace \midrule[0.1pt]\addlinespace
       $\ket{\downarrow_2}$  & $\frac{  \ket{n_2}+ \ee^{\ii 2 \vartheta_2}  \ket{n_2+2}}{\sqrt{2}} $ & $\frac{\ee^{\ii \vartheta_{2}/2} \ket{n_2}+\ee^{ -\ii \vartheta_{2}/2} \ket{n_2-1}}{\sqrt{2}}$ \\ \addlinespace
       $\ket{\uparrow_2}$  & $\frac{\ket{n_2-1}+ \ee^{\ii  2 \vartheta_2}  \ket{n_2+1}}{\sqrt{2}}$ & $\frac{ \ee^{\ii 3\vartheta_{2}/2} \ket{n_2+1}+\ee^{- \ii  3\vartheta_{2}/2} \ket{n_2-2}}{\sqrt{2}}$ \\\addlinespace
       \bottomrule[0.1pt] \bottomrule[0.1pt]
    \end{tabularx}
    \label{tab:subset}
\end{table}
In the following discussion, we restrict ourselves to the regime of high photon numbers.
As before, we introduce the GHZ state
\begin{align}\label{eq:ghz_superpos}
    \ket{\mathrm{GHZ}} = \frac{1}{\sqrt{2}} \qty( \ee^{\ii \vartheta/2} \ket{\downarrow_0,\downarrow_1,\downarrow_2} + \ee^{-\ii \vartheta/2 } \ket{\uparrow_0,\uparrow_1,\uparrow_2}).
\end{align}
Similar to Fock states, only certain choices of photon numbers lead to a nonvanishing interference signal.
One specific choice is listed in Table~\ref{tab:subset}.
In the table, we also made a specific choice for the phases $\vartheta_i$ of the superposition, which are connected to $\vartheta$ through the relation $\vartheta=\vartheta_0 + \vartheta_1 + \vartheta_2$.
The resulting interferometer phase corresponds to the one from Eq.~\eqref{eq:interferometerphase}, and the visibility takes the value
\begin{equation}
    \mathcal{V} = 9/16 > 1/2.
\end{equation}
As expected, initially entangled states reach a higher visibility once we replace the Fock states with suitable superpositions.

For the W state consisting of Fock states, we did not observe any interference because full welcher-Weg information was still encoded into the light fields.
However, for
\begin{align}
    \ket{W} = \frac{1}{\sqrt{3}} \qty( \ket{\downarrow_0,\downarrow_1,\uparrow_2} + \ket{\downarrow_0,\uparrow_1,\downarrow_2} + \ket{\uparrow_0,\downarrow_1,\downarrow_2})
\end{align}
involving superpositions of Fock states from Eq.~\eqref{eq:basis_superposition}, we can overcome this obstacle. 
With a suitable choice of photon numbers and phases listed in Table~\ref{tab:subset}, we find the interferometer phase from Eq.~\eqref{eq:interferometerphase}, where again the phases $\vartheta_i$ from Table~\ref{tab:subset} define $\vartheta = \vartheta_0 + \vartheta_1 + \vartheta_2$. 
The visibility for this subset of states resolves to 
\begin{equation}
    \mathcal{V} = 1/4< 9/16.
\end{equation}
In contrast to Fock states, the superpositions are able to partially erase welcher-Weg information that was encoded into the light fields before.
However, the maximum visibility for entangled superposed Fock states is still larger for GHZ states than for W states.

To gain more insight into the behavior of the visibility under the influence of both entanglement and superpositions, we first use the GHZ state with Fock numbers as defined in Table~\ref{tab:subset} and find the blue dashed line when varying $\alpha$ and $\beta$.
We observe again a square-root-like behavior similiar to Fock states.
However, since we use a superposition as a two-state subset, we arrive in the limit $\tau_3 \to 0$ at a separable state that is a superposition, which has $\mathcal{V}=1/8$ as already shown in Ref.~\cite{Soukup2021}.
This limit can be motivated as follows:
The choice of photon numbers similar to Eq.~\eqref{eq:ghz-like} with entangled Fock states yields a nonvanishing overlap.
An even superposition in each of the three fields implies that only one-half of each field contributes to the interference signal and thus it decreases by $(1/2)^3=1/8$.
Similar to the discussion above, we expand the GHZ state by adding more components and choosing the same equal superpositions to study the 2-1 and 2-2 state. 
If we use in the subset of Eq.~\eqref{eq:basis_superposition} the photon numbers of the GHZ state from Table~\ref{tab:subset}, the visibility of the dashed curve cannot be enhanced as the additional components will vanish in the overlap trivially.
Thus, we improve the visibility by selecting suitable Fock numbers to reduce the amount of vanishing components in the overlap. 
Using this strategy, we find the visibilities for the 2-1 and 2-2 state in Fig.~\ref{fig:result} that are marked by crosses together with the GHZ and W state.

In addition, the 2-3 state from the table of Fig.~\ref{fig:entanglement} is a generalization of all the other discussed states.
Similar to the discussion above, we now continuously vary  $\tau_3$ by scanning the coefficients given in the table of Fig.~\ref{fig:entanglement}.
As mentioned before, for each value of $\tau_3$ we can enhance the visibility by an optimal choice of Fock numbers of the subsets defined in Eq.~\eqref{eq:basis_superposition}.
The resulting solid blue curve in Fig.~\ref{fig:result} denotes the upper bound \footnote{These results were obtained in the high-photon limit, where we replace the trigonometric functions by $1/\sqrt{2}$ and unity, respectively.}. 
The discontinuous parts originate from points where different Fock numbers yield better visibilities.
While the GHZ state again stands out with the best visibility ($\tau_3=1$), once we involve superpositions in the two-state subset, we do not necessarily observe a continuous decrease of the visibility ($\tau_3<1$) anymore due to the competing effects of entanglement and superpositions that both diminish welcher-Weg information.  

\section{Conclusions}
We have determined the interference signal of a Mach-Zehnder interferometer where quantized light fields serve as atom-optical elements.
Within a two-state subset of possible states, we used $\tau_3$ to quantify tripartite entanglement.
In fact, the loss in visibility due to welcher-Weg information can be erased partially by resorting to initially entangled light fields.
In the simplest case of Fock states in the two-state subset and a GHZ state, initial entanglement can be used to prevent the encoding of complete welcher-Weg information that suppresses the interference signal.
We were able to show that a visibility of $\mathcal{V}=1/2$ is possible in the high-photon limit.
We found an analytical dependence of the visibility $\mathcal{V} \propto \sqrt{\tau_3}$ on the amount of residual entanglement for all investigated initial states.
In addition, we already knew from previous works that involving separable superpositions yields an improvement in the visibility.
Consequently, we were able to demonstrate that a combination of both initial entaglement and superpositions leads to the best interference signal for our choice of possible states.
Indeed, our results underline that it is not mere entanglement or superpositions that cause an increase in visibility, but in fact the amount of welcher-Weg information encoded into the light fields.

Even though cavities with high finesse can drive Rabi cycles with low photon numbers, they are impractical for atom interferometry, where large separations and therefore large mode volumes are required.
Nonetheless, cavity-based schemes are currently explored for atom interferometry~\cite{Hamilton2015,Canuel2018,Xu2019,Nourshagh2020, Nourshagh2020_2}, but they do not target a regime where quantization effects are observable.
However, realizing three entangled fields adds another level of complexity.
Entangling three spatially separated modes or introducing a temporal delay comes with severe issues for the experimental implementation, not covered in our article.
Moreover, single-photon transitions limit the duration of the experiment to the lifetime of the excited state.
In conventional interferometers, this limitation is circumvented by the help of Bragg- or Raman-based traveling-wave setups.
In principal, our treatment could be generalized to such two-photon transitions.

Our article highlights that entangled light fields can be used to mitigate some of the deleterious effects of field quantization.
Still, perfect visibility, and by that, optimal sensitivity, is obtained for classical fields or sufficiently strong coherent states.
Hence, there is no direct benefit from working in the considered regime. 
However, we only discussed a first-quantized matter wave.
The nonlinear interaction and entanglement of second-quantized many-particle atomic and optical fields are expected to enhance the interferometric sensitivity beyond the shot-noise limit~\cite{Davis2016, Hosten2016, Haine2013, Haine2015, Haine2016, Greve2021}.
Thus, we expect further studies of quantized beam splitters and mirrors acting on second-quantized atomic systems to demonstrate a true metrological gain.

The quantization of the diffracting light fields is still irrelevant in today's devices with intense laser pulses.
However, the evolution of the field towards cavity-based atom interferometers is a first step towards regimes where such quantization effects and the encoding of which-Way information may become non-negligible.
Our results highlight new leverage points to circumvent these problems and once more emphasize the role of complementarity in interferometric experiments.

\begin{acknowledgements}
We are grateful to W. P. Schleich for his stimulating input and continuing support.
Moreover, we thank A. Friedrich, C. Pfleghar, K. Soukup, and C. Ufrecht, as well as the QUANTUS and INTENTAS teams, for fruitful and interesting discussions.
The QUANTUS and INTENTAS projects are supported by the German Aerospace Center (Deutsches Zentrum f\"ur Luft- und Raumfahrt, DLR) with funds provided by the Federal Ministry of Economic Affairs and Energy (Bundesministerium f\"ur Wirtschaft und Energie, BMWi) due to an enactment of the German Bundestag under Grant No. 50WM1956 (QUANTUS V), No. 50WM2177, and 50WM2178 (INTENTAS).
The projects ``Building composite particles from quantum field theory on dilaton gravity'' (BOnD) and ``Metrology with interfering Unruh-DeWitt detectors'' (MIUnD) are funded by the Carl Zeiss Foundation (Carl-Zeiss-Stiftung).
The work of IQ\textsuperscript{ST} is financially supported by the Ministry of Science, Research and Art Baden-W\"urttemberg (Ministerium f\"ur Wissenschaft, Forschung und Kunst Baden-W\"urttemberg).
E.G. thanks the German Research Foundation (Deutsche Forschungsgemeinschaft, DFG) for a Mercator Fellowship within CRC 1227 (DQ-mat).
\end{acknowledgements}

\appendix

\section{Branch-dependent operators}
\label{sec:appendix}
Assuming that the atom enters the interferometer in the ground state $\ket{g}$, the interference signal observed in the exit port defined by the ground state
\begin{align}
    I = \bra{\Psi}\otimes \bra{g} \hat{U}_{\mathrm{MZ}}^\dagger \hat{\Pi}_g \hat{U}_{\mathrm{MZ}} \ket{g} \otimes \ket{\Psi} 
\end{align}
arises from the expectation value of the projection operator $\hat{\Pi}_g= \ket{g}\bra{g}$.
Hence, the interferometer can be effectively described by the postselected operator $\hat{\mathcal{O}}\equiv \bra{g} \hat{U}_{\mathrm{MZ}} \ket{g}$ that only acts on the c.m. motion as well as the light fields. 
The scattering operators $\hat{\mathcal{S}}^{\qty(\ell)}$ have a matrix representation with elements $\hat{\mathcal{S}}^{\qty(\ell)}_{ij} = \bra{i} \hat{\mathcal{S}}^{\qty(\ell)} \ket{j}$ according to Eq.~\eqref{eq:ScattMatrix} with $i,j \in \lbrace g,e \rbrace$. Similarly, the evolution from Eq.~\eqref{eq:U} is diagonal
\begin{align}
    \hat{U}= \exp{-\ii\left(\frac{\hat{p}^2}{2m\hbar}+ \sum_{\ell=0}^2{ \omega \hat{n}_\ell} \right)T} \qty( \op{g} + \ee^{-\ii \omega_a T}  \op{e} ),
\end{align}
and its matrix elements can be defined analogously.
Switching to this matrix representation, the effective operator is determined by 
\begin{align}
    \hat{\mathcal{O}} &= \bra{g} \hat{\mathcal{S}}^{\qty(2)}\hat{U}\hat{\mathcal{S}}^{\qty(1)}\hat{U}\hat{\mathcal{S}}^{\qty(0)} \ket{g} \\ \notag
    &= \begin{pmatrix} 1 \\ 0 \end{pmatrix}^\mathrm{T} \begin{pmatrix} \hat{\mathcal{S}}^{\qty(2)}_{gg} & \hat{\mathcal{S}}^{\qty(2)}_{ge} \\ \hat{\mathcal{S}}^{\qty(2)}_{eg} & \hat{\mathcal{S}}^{\qty(2)}_{ee}  \end{pmatrix} \begin{pmatrix} \hat{U}_{gg} & 0 \\ 0 & \hat{U}_{ee} \end{pmatrix} \begin{pmatrix} \hat{\mathcal{S}}^{\qty(1)}_{gg} & \hat{\mathcal{S}}^{\qty(1)}_{ge} \\ \hat{\mathcal{S}}^{\qty(1)}_{eg} & \hat{\mathcal{S}}^{\qty(1)}_{ee}  \end{pmatrix}  \\
    & \hspace{1cm}\times \begin{pmatrix} \hat{U}_{gg} & 0 \\ 0 & \hat{U}_{ee} \end{pmatrix} \begin{pmatrix} \hat{\mathcal{S}}^{\qty(0)}_{gg} & \hat{\mathcal{S}}^{\qty(0)}_{ge} \\ \hat{\mathcal{S}}^{\qty(0)}_{eg} & \hat{\mathcal{S}}^{\qty(0)}_{ee}  \end{pmatrix} \begin{pmatrix} 1 \\ 0 \end{pmatrix}.
\end{align}
After the matrix multiplication, we arrive at four possible paths
\begin{align}
\begin{split}
    \hat{\mathcal{O}} &= \hat{\mathcal{S}}^{\qty(2)}_{ge} \hat{U}_{ee} \hat{\mathcal{S}}^{\qty(1)}_{eg}\hat{U}_{gg}\hat{\mathcal{S}}^{\qty(0)}_{gg} + \hat{\mathcal{S}}^{\qty(2)}_{gg}\hat{U}_{gg}\hat{\mathcal{S}}^{\qty(1)}_{ge}\hat{U}_{ee}\hat{\mathcal{S}}^{\qty(0)}_{eg} \\ \label{eq:multiplication}
    &+ \hat{\mathcal{S}}^{\qty(2)}_{gg}\hat{U}_{gg} \hat{\mathcal{S}}^{\qty(1)}_{gg}\hat{U}_{gg}\hat{\mathcal{S}}^{\qty(0)}_{gg} +\hat{\mathcal{S}}^{\qty(2)}_{ge}\hat{U}_{ee} \hat{\mathcal{S}}^{\qty(1)}_{ee}\hat{U}_{ee}\hat{\mathcal{S}}^{\qty(0)}_{eg}
\end{split}
\end{align}
through the interferometer. 
However, the two paths in the second line of  Eq.~\eqref{eq:multiplication} correspond to the dotted lines in Fig.~\ref{fig:1} and will be spatially separated from the other two paths.
As a consequence, they do not contribute to the interference signal and can be disregarded, which is the origin of the atom loss ocurring during the mirror pulse.
This procedure can be formalized by a postselection on a spatial region.
Comparing the remaining two paths through the interferometer in Eq.~\eqref{eq:multiplication} with Fig.~\ref{fig:1}, we identify
\begin{subequations}
\begin{align}
     \hat{\mathcal{O}}_l\equiv \hat{\mathcal{S}}^{\qty(2)}_{ge} \hat{U}_{ee} \hat{\mathcal{S}}^{\qty(1)}_{eg}\hat{U}_{gg}\hat{\mathcal{S}}^{\qty(0)}_{gg}
\end{align}
as the operator describing the motion along the lower path, and the operator
\begin{align}
     \hat{\mathcal{O}}_u \equiv \hat{\mathcal{S}}^{\qty(2)}_{gg}\hat{U}_{gg}\hat{\mathcal{S}}^{\qty(1)}_{ge}\hat{U}_{ee}\hat{\mathcal{S}}^{\qty(0)}_{eg}
\end{align}
\end{subequations}
is the propagation along the upper path.

\bibliography{Literatur}

\end{document}